\documentclass[a4paper,prd,superscriptaddress,twocolumn,aps,reprint,notitlepage,floatfix]{revtex4-1}
\usepackage{amssymb,amsmath}
\usepackage[normalem]{ulem}
\usepackage[pdftex]{graphicx}
\usepackage[usenames]{color}
\definecolor{darkred}{rgb}{0.7,0,0}
\definecolor{darkgreen}{rgb}{0,0.5,0}
\definecolor{darkblue}{rgb}{0,0,0.7}
\usepackage[pdftex, unicode, colorlinks, bookmarks=true, citecolor=darkblue, link color=darkblue, urlcolor=darkblue]{hyperref}

\newcommand{\abs}[1]{\left|#1\right|}
\newcommand{\ii}{\mathrm{i}}

\begin{document}


\title{Trade-off between quantum and thermal fluctuations in mirror coatings yields improved sensitivity of gravitational-wave interferometers}


\author{N.V.~Voronchev}
\email{n.voronchev@physics.msu.ru}
\affiliation{Faculty of Physics, Moscow State University, Moscow 119991, Russia}
\author{S.L.~Danilishin}
\email{shtefan.danilishin@uwa.edu.au}
\affiliation{School of Physics, University of Western Australia, 35 Stirling Hwy, Crawley 6009, WA, Australia}
\author{F.Ya.~Khalili}
\email{khalili@phys.msu.ru}
\affiliation{Faculty of Physics, Moscow State University, Moscow 119991, Russia}

\begin{abstract}

We propose a simple way to improve the laser gravitational-wave detectors sensitivity by means of reduction of the number of reflective coating layers of the core optics mirrors. This effects in the proportional decrease of the coating thermal noise, the most notorious among the interferometers technical noise sources. The price for this is the increased quantum noise, as well as high requirements for the pump laser power and power at the beamsplitter. However, as far as these processes depend differently on the coating thickness, we demonstrate that a certain trade-off is possible, yielding a 20-30\% gain (for diverse gravitational wave signal types and interferometer configurations), providing that feasible values of laser power and power on the beamsplitter are assumed.

\end{abstract}

\maketitle

\section{Introduction}

Among the multitude of noise sources limiting the sensitivity of contemporary laser gravitational-wave (GW) detectors (LIGO \cite{LIGOsite, Waldman2006}, VIRGO \cite{VIRGOsite, Acernese2006}, GEO-600 \cite{GEOsite, Hild2006}, and TAMA \cite{TAMAsite, Ando2005}), the one, usually referred to as \emph{quantum} noise and stemming from the quantum nature of light, stands apart from the rest of the noise sources, referred to as \emph{technical} or \emph{classical} ones, respectively. Quantum noise originates from quantum fluctuations of phase (shot noise) and intensity (radiation-pressure noise) of light circulating inside the interferometers, which obey Heisenberg's uncertainty relation \cite{Caves1981} and therefore can not be reduced simultaneously. The latter group comprises various fluctuations of thermal, seismic and similar origin that can be, in principle, diminished either by cooling, or using better materials, more sophisticated seismic isolation and so on. All the hitherto undertaken efforts towards the improvement of GW interferometers sensitivity went in two parallel but virtually independent streams. In many proposed methods of diverse technical noise sources mitigation \cite{Mours2006, Bondaresku2006, Harry2007, Kimble_PRL_101_260602_2008, Bondaresku2008, Villar2010, Hong2011, 11a1KoGuGo, HarryBook2012}, the authors assumed quantum noise of the interferometer as independent of the technical noise budget and thus did not take it into consideration. Equally lukewarm were the researchers of the quantum noise, who proposed a plenty of sophisticated and witty ways for its reduction \cite{Unruh1982, 90a1BrKh, 99a1BrKh, Buonanno2001, 01a2Kh, Buonanno2002, 02a1KiLeMaThVy, Chen2002, Corbitt2004-3, 04a1Da, Rehbein2008, 08a1KoSiKhDa, 09a1KhMiCh, 09a1ChDaKhMu, 11a1KhDaMuMiChZh}, towards the technical noise sources, tacitly implying them being independent on the quantum fluctuations of light inside the interferometer.

However, the growing interest to the optimal configurations of future GW detectors, inspired by the recent achievements in reduction of technical noise in first-generation detectors and a proximity of the start of construction of the second generation ones, has brought the problem of simultaneous treatment of the two groups of noise sources to the fore \cite{Rehbein2008, 08a1KoSiKhDa}. Indeed, the sensitivity of the current first generation detectors (the most sensitive of them, LIGO, has already finished its life cycle) is limited mostly by seismic noise at lower frequencies (below $\sim100\,{\rm Hz}$) and by the quantum shot noise at higher frequencies. In the next-generation detectors, such as Advanced LIGO  \cite{AdvLIGOsite, Harry2010}, Advanced VIRGO \cite{AdvVIRGOsite}, and LCGT \cite{LCGTsite}, the technical noise will be reduced significantly by using much better seismic isolation and other technological advances. Along with this, the quantum shot noise will be suppressed by about one order of magnitude due to increased optical power and, very probably, by the injection of the quantum squeezed light into the interferometer, as it has been proposed first in \cite{Caves1981} (this technology has been successfully tested experimentally in GEO-600 \cite{Nature_2011}). Yet the second generation detectors sensitivity remain bound by the mix of the technical and quantum noise: shot noise at high frequencies, (ii) radiation-pressure noise at low frequencies, and (iii) coatings thermal noise in the best sensitivity medium frequencies band around $100\,{\rm Hz}$ (see Fig.\,3 of the paper \cite{Harry2010}).

In this article, we make a further step forward and consider quantum noise and coating thermal noise in conjunction. The importance of thermal fluctuations in core optics dielectric coatings was realised by the community several years ago \cite{Harry2006}. All the hitherto proposed ways to reduce these fluctuations, were it based on broadening of laser beams \cite{Mours2006, Bondaresku2006, Bondaresku2008, Hong2011}, or on using better coating materials \cite{Harry2007}, or on coating structure optimisation \cite{Kimble_PRL_101_260602_2008, Villar2010, 11a1KoGuGo},  followed implicitly a common rule that the end mirrors (ETMs) of the interferometer should be as reflective as possible thus requiring the number of coating layers to be pretty large (ca. 40). But since the power spectral density of these fluctuation rises linearly with the number of coating layers, the improvement provided by these methods is rather modest.

Another apparent way to get rid of coating noise by getting rid of the (at least a part of) the coating itself was proposed even earlier. In \cite{04a1BrVy}, authors suggested to replace the end mirrors by coatingless corner reflectors, while in \cite{05a1Kh} the short anti-resonance-tuned Fabry-P\'erot cavities have to play the role of ETMs. Even more radical solution to use the pass-through Mach-Zehnder/Fabry-P\'erot topology instead of the Michelson/Fabry-P\'erot one was suggested in \cite{10a2Kh}. However, the implementation of these methods in the near future is improbable, for corner reflectors were shown to have high optical losses \cite{05a1Ta} and the latter two solutions require too radical modifications of the GW detector optical setup.

\begin{figure*}
  \includegraphics[width = 0.45\textwidth]{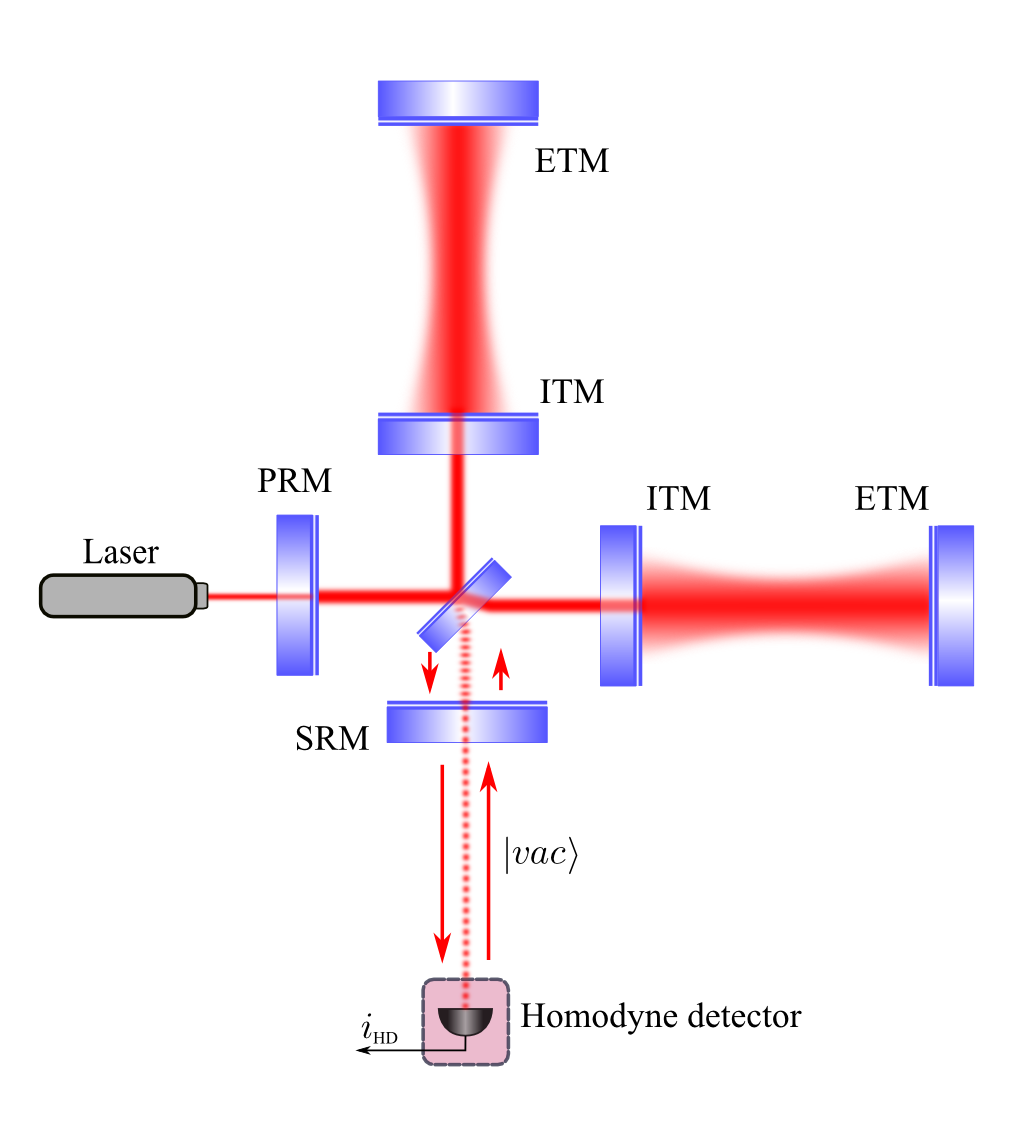} \hfill
  \includegraphics[width = 0.45\textwidth]{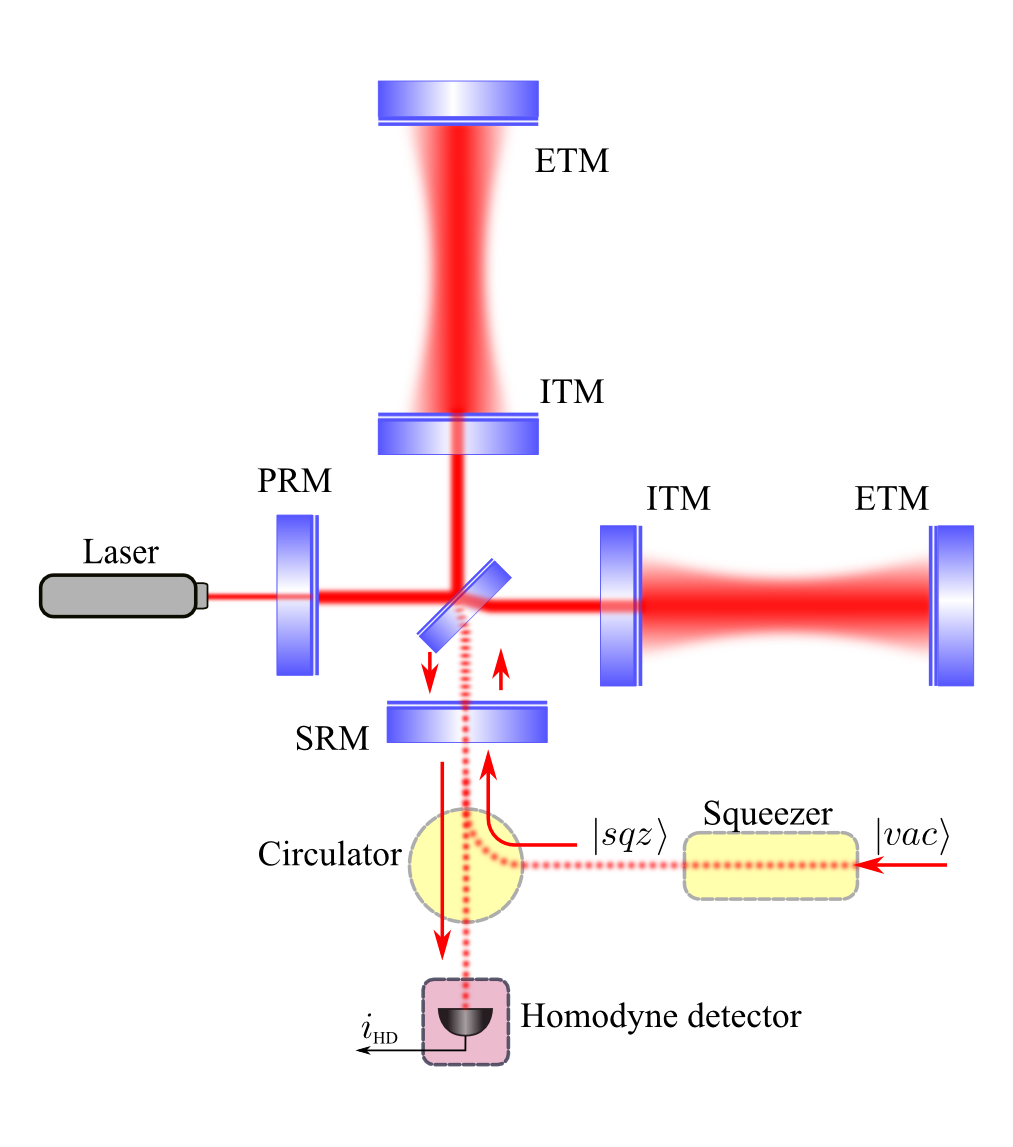}   \\ \vspace{1em}
  \includegraphics[width = 0.45\textwidth]{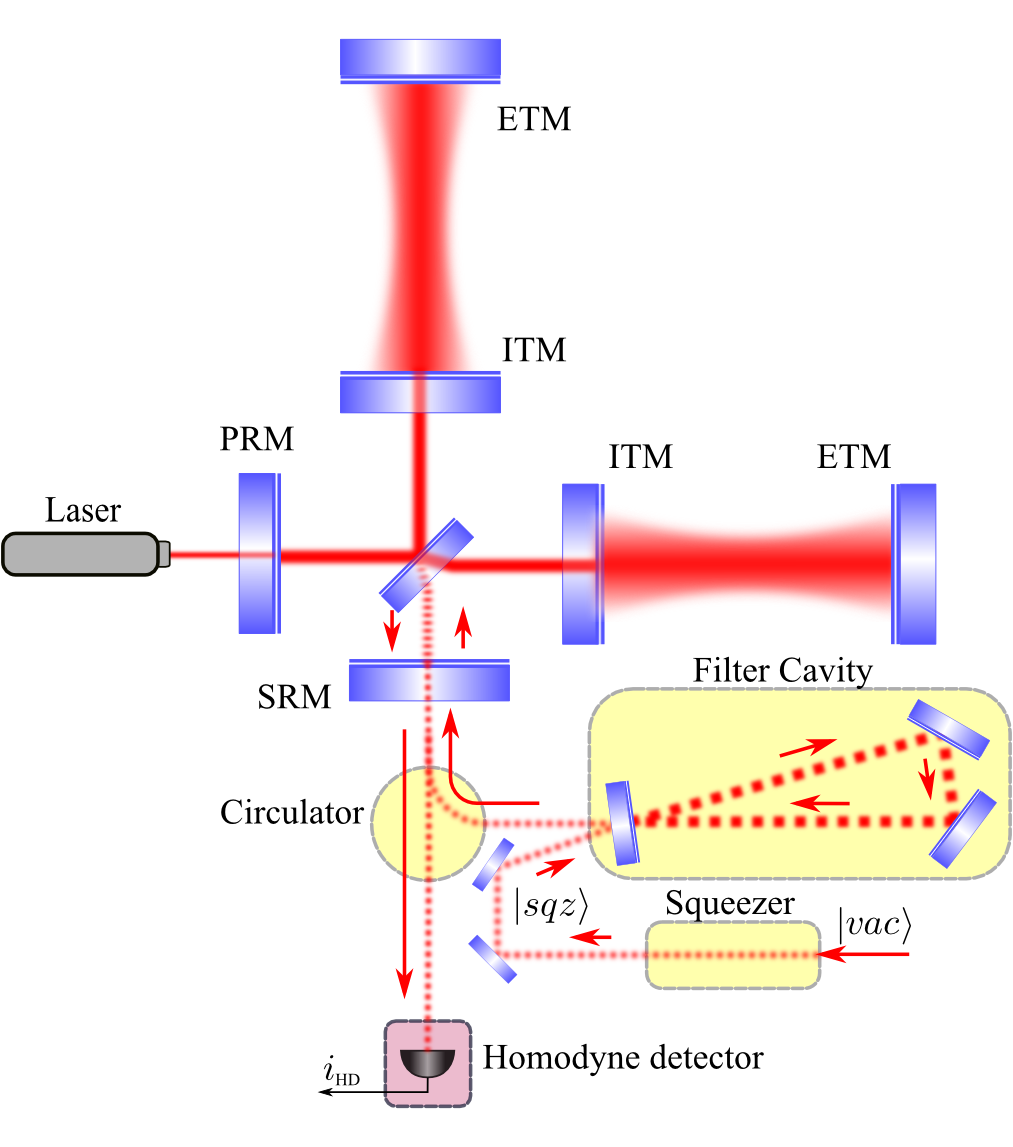}  \hfill
  \includegraphics[width = 0.45\textwidth]{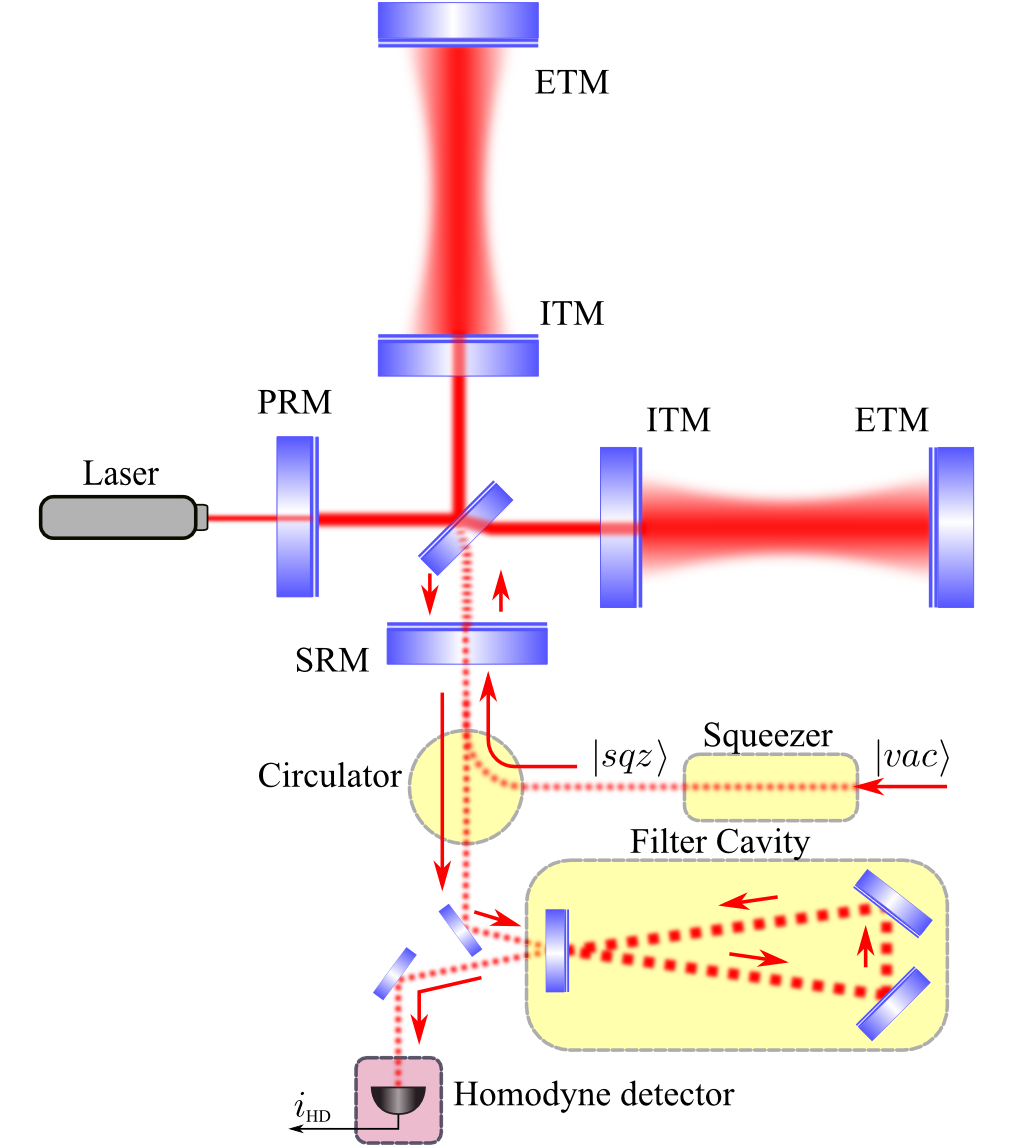}
  \caption{The schemes of the advanced gravitation-wave detector being considered in this paper. Top left: the signal- and power-recycled interferometer similar to the one planned for Advanced LIGO \cite{Harry2010} (referred to as ``plain''). Top right: the configuration with frequency-independent squeezed light injection into the dark port (referred to as ``squeezed''). Bottom left: the configuration with injection of frequency-dependent squeezed light created by means of the input filter cavity (referred to as ``pre-filtering''). Bottom right: the back action evading configuration with the additional output filter cavity and the squeezed light injection into the dark port (referred to as ``post-filtering'').}
  \label{fig:detector}
\end{figure*}

The requirement for the ETMs of the standard Michelson/Fabry-P\'erot scheme of GW interferometer (see Fig.~\ref{fig:detector}) to have high reflectivity has rather strong logic behind. There are two obvious reasons in favour of it. First, in the power-recycled topology, the value of ETM power reflectivity $R_{\rm ETM}$ defines how much circulating power $I_c$ can be built up in the arm cavities, for a given value of the input laser power $I_0$:
\begin{equation}
  I_c \le \frac{I_0}{2(1-R_{\rm ETM})} \,.
\end{equation}
Simple estimate based on the values of $I_0$ and and $I_c$ planned for the Advanced LIGO yields $1-R_{\rm ETM}\lesssim10^{-4}$. Second, a non-ideal reflectivity of the end mirrors means an injection of additional optical vacuum fluctuations into the arm cavities, that is, the increase of the quantum noise.

But if we ask ourselves a question whether this requirement always provides unconditionally optimal sensitivity for the GW interferometer, we claim that the answer will be `no'. The circulating power issue can be solved either by using more powerful laser, as {\it e.g.} the one with $I_0 = 500$~W proposed for the third generation Einstein Telescope gravitation-wave detector \cite{Sathyaprakash2011}, or by using squeezed vacuum injection as it might be done in Advanced LIGO \cite{Adhikari_private}.

Regarding the influence of the additional vacuum entering the arm cavities through the more transparent ETMs, it is evident, that in the scenario where the total noise budget, in the best sensitivity frequency band, is dominated by the mirrors coatings thermal noise, it is reasonable to reduce the number of coating layers of the end mirrors, increasing the quantum noise, but decreasing the coatings thermal noise.

Therefore, the number of layers of the core optics coatings should be included into the set of optical parameters, such as arm cavities bandwidth, the signal recycling mirror transmittance and the the signal recycling cavity detuning, over which the minimisation of the sum noise of the interferometer is run.

In this paper, we perform this kind of optimisation for the signal- and power-recycled Fabry-P{\'e}rot--Michelson interferometer (see Fig.\,\ref{fig:detector}), assuming its main parameters close to the ones planned for the Advanced LIGO. In the next section, we describe the variants of the advanced gravitation-wave detectors scheme which we optimise, the model of technical noise and quantum noise which we use, and the optimisation procedure. In Sec.\,\ref{sec:conclusion}, we discuss the results of optimisation.

\section{Optimised configurations and the optimisation procedure}

\subsection{Interferometer configurations}

The following configurations are considered in this paper (see Fig.\,\ref{fig:detector}):

\begin{itemize}
  \item Plain: the ordinary signal- and power-recycled interferometer, similar to the Advanced LIGO, with vacuum input (no squeezing).
  \item Squeezed: the same as the above, but with squeezed light injection into the dark port.
  \item Pre-filtering: the same as previous, but with frequency-dependent squeezing angle implemented by means of the additional input filter cavity.
  \item Post-filtering: the back action evading configuration with the additional output filter cavity and the squeezed light injection into the dark port.
\end{itemize}

The last two configurations, in a more sophisticated two-cavities form, were first proposed in paper \cite{02a1KiLeMaThVy}. Here we consider more simple cases with only one relatively short filter cavity. Estimates (see, {\it e.g}, \cite{10a1Kh}) show that in the presence of the technical noise, single filter cavity configurations provide almost the same sensitivity as the two-cavity ones, while being much less expensive in implementation. It was shown also in the paper \cite{10a1Kh} that the sensitivity of these {\it phase filtering} schemes is better than that of the {\it amplitude filtering} configuration \cite{Corbitt2004-3, 07a1Kh, 09a1KhMiCh}. Currently, it is these two schemes that are considered as the most probable variants for upgrading the Advanced LIGO.

\begin{table}
  \caption{The main parameters and their numerical values.}\label{tab:params}
  \begin{ruledtabular}
    \begin{tabular}{ccl}
      Parameter\!\!\!& Value                               & Description \\
      \hline
      $\omega_p$    & $2\pi c/(1.064\,\mu{\rm m})$        & Optical pump frequency \\
      $M$           & $40\,{\rm kg}$                      & Mirrors mass \\
      $L$           & $4\,{\rm km}$                       & Interferometer arms length \\
      $I_0$         & $\le500\,{\rm W}$                   & Input optical power \\
      $I_c$         & $200\,{\rm kW}$ or $840\,{\rm kW}$  & Circulating power in the arms\\
      $I_{\rm abs}$ & $\le 1\,{\rm W}$                    & Power absorption in the ITMs\\
        & & and in the beamsplitter. \\
      $\eta$        & 0.9                                 & Photodetectors quantum \\
        & & efficiency \\
      $l$           & $50\,{\rm m}$                       & Filter cavity length \\
      $A_f$         & $10\,{\rm ppm}$                     & Filter cavity losses per bounce \\
    \end{tabular}
  \end{ruledtabular}
\end{table}

For the main parameters of the considered schemes, we assume the values close to the ones planned for the Advanced LIGO and its upgrades, see Table\,\ref{tab:params}. We suppose, however, that the input power can be higher, up to $500\,{\rm W}$, and for the circulating power, along with the ``canonical'' value of $840\,{\rm kW}$, we consider the reduced one equal to $200\,{\rm kW}$.

\subsection{Figures of merit}

The common figure of merit for the gravitational wave detectors sensitivity is the signal-to-noise ratio integral (SNR) for some standard gravitational waves source:
\begin{equation}\label{SNR_BNS}
  \rho^2({\bf p},N_{\rm ITM},N_{\rm ETM}) = \int_{\Omega_{\rm min}}^{\Omega_{\rm max}}\!\!
    \frac{|h(\Omega)|^2}{S^h(\Omega;N_{\rm ITM},N_{\rm ETM};{\bf p})}
    \frac{d\Omega}{2\pi} \,,
\end{equation}
where $\Omega_{\rm min}$ and $\Omega_{\rm max}$ are the minimal and the maximal frequencies of the GW detector sensitivity band, which we assume to be equal to  $2\pi\times5\,{\rm Hz}$ and $2\pi\times5\,{\rm kHz}$, respectively; $h(\Omega)$ is the gravitational wave strain signal spectrum;
\begin{multline}
  S^h(\Omega;N_{\rm ITM},N_{\rm ETM};{\bf p})
  = S^h_{\rm quant}(\Omega;N_{\rm ITM},N_{\rm ETM};{\bf p}) \\
    + S^h_{\rm coat}(\Omega;N_{\rm ITM},N_{\rm ETM}) + S^h_{\rm tech}(\Omega)
\end{multline}
is the spectral density of the sum noise normalized as the equivalent gravitation wave strain variation; $S^h_{\rm quant}(\Omega;N_{\rm ITM},N_{\rm ETM};{\bf p})$ is the quantum noise spectral density which depends on the numbers $N_{\rm ITM}$ and $N_{\rm ETM}$ of the coating layers doublets of the input and the end mirrors, respectively, as well as on the vector ${\bf p}$ of other optical parameters of the interferometer (we  specify this vector below); $S^h_{\rm coat}(\Omega;N_{\rm ITM},N_{\rm ETM})$ is the mirrors coating thermal noise spectral density, which also depends on $N_{\rm ITM}$ and $N_{\rm ETM}$; and $S^h_{\rm tech}(\Omega)$ is the sum spectral density of the rest of technical noise sources.

For our optimisation, we will use two standard types of the gravitational waves sources, (see, {\it e.g}, \cite{Postnov2006}). The first one is the GW signal from the inspiral stage of the binary neutron stars (BNS) collisions with
\begin{equation}\label{SNR_Burst}
  |h_{\rm BNS}(\Omega)|^2 = K_{\rm BNS}\times\begin{cases}
    \Omega^{-7/3} \,, & \Omega\le2\pi\times1.5\,{\rm kHz}\,,\\
    0 \,, & \Omega>2\pi\times1.5\,{\rm kHz} \,,
  \end{cases}
\end{equation}
which accounts mostly for the low-frequency noise. The second one is the gravitational wave ``bursts'', with
\begin{equation}
  |h_{\rm Burst}(\Omega)|^2 = \frac{K_{\rm Burst}}{\Omega} \,,
\end{equation}
which require more broadband sensitivity from the detector. Here $K_{\rm BNS}$ and $K_{\rm Burst}$ are some factors, depending on the astrophysical parameters of the signal source but not on the observation frequency $\Omega$ and detector optical parameters.

Our treatment of the interferometer quantum noise is based on the works \cite{02a1KiLeMaThVy, Buonanno2003, 10a1Kh}. The corresponding explicit equations for the quantum noise spectral densities $S^h_{\rm quant}$ are too cumbersome to be shown here, especially for the filter cavities based configurations, and can be found in Appendix \ref{app:quantum}.

In our calculation of the coating Brownian noise spectral density $S^h_{\rm coat}$, we follow the paper \cite{11a1KoGuGo}, which takes into account additional effects of light interference in the coating layers as well as the photoelastic effect. This method gives more precise correct estimate for the coating Brownian noise spectral density, which is smaller than the previous estimates \cite{Harry2007} by 3\% - 13\%, depending on coating layers number.

For the value of spectral density $S^h_{\rm tech}$ of the other kinds of technical noise, we rely on the GWINC software tool \cite{GWINCsite}.

\subsection{Optimisation procedure}

For each of the four configurations, of the two figures of merit described above, and of the two values of circulating power, $I_c=840\,{\rm kW}$ and $I_c=200\,{\rm kW}$ (16 variants total), we found maximums of the SNR integrals \eqref{SNR_BNS} and \eqref{SNR_Burst} over the parameter vector ${\bf p}$ and in the numbers of coating layers doublets of ITM, $N_{\rm ITM}$, and ETM, $N_{\rm ETM}$.

This optimisation was performed in two steps. First, we maximised the SNRs in ${\bf p}$ for each of $N_{\rm ITM}$ and $N_{\rm ETM}$, using  Nelder-Mead simplex method \cite{GSLsite}, and getting thus the semi-optimised values of SNR $\rho^2(N_{\rm ITM}, N_{\rm ETM})$ as a function of the numbers of coating layers doublets. At the second step, we found the optimal values of $N_{\rm ITM}$ and $N_{\rm ETM}$ by a simple grid search.

The following interferometer parameters were included into the parameter space of the vector ${\bf p}$ (depending on the configuration):
\begin{subequations}\label{bf_p}
  \begin{align}
    \text{Plain:} & \ {\bf p} = \{R_{\rm SRM}, \phi_{\rm SR}, \phi_{\rm LO}\} , \\
    \text{Squeezed:} &
        \ {\bf p} = \{R_{\rm SRM}, \phi_{\rm SR}, \phi_{\rm LO}, e^r, \lambda\} , \\
    \text{Pre-/post-filtering:} &
      \ {\bf p}
        = \{R_{\rm SRM}, \phi_{\rm SR}, \phi_{\rm LO}, e^r, \lambda, \gamma_f, \delta_f\},
  \end{align}
\end{subequations}
where $R_{\rm SRM}$ is the signal recycling mirror power reflectivity, $\phi_{\rm SR}$ is the signal recycling cavity detuning angle, $\phi_{\rm LO}$ is the homodyne angle, $e^r$ is the squeezing factor and $\lambda$ is the squeezing angle, $\gamma_f$ is the filter cavity half-bandwidth and $\delta_f$ is the filter cavity detuning. We limited the  squeezing factor by 10\,dB ($e^r\le\sqrt{10}$), according to the contemporary experimental achievements in the low-frequency squeezing \cite{Vahlbruch_PRL_100_033602_2008, Mehmet2011}. The input laser power was limited by $500\,{\rm W}$.

\begin{figure}
  \includegraphics[width = 0.4\textwidth]{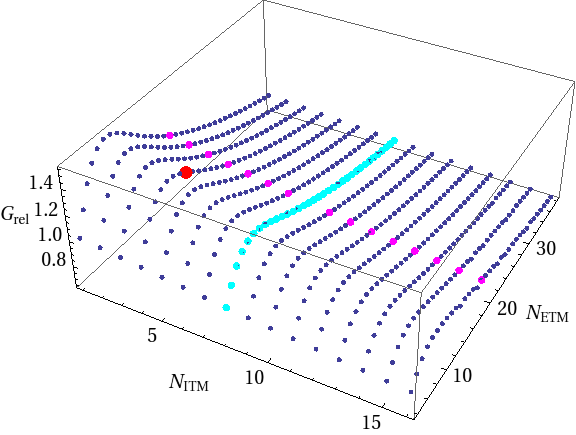}
  \caption{Relative gain in SNR as a function of the coating layers doublets numbers $N_{\rm ITM}, N_{\rm ETM}$. Cyan and magenta dots correspond to the default numbers $N_{\rm ITM}^{\rm def}$, $N_{\rm ETM}^{\rm def}$ \eqref{N_def} given by GWINC.}
  \label{fig:SNR-NumLayers-3D}
\end{figure}

In Fig.~\ref{fig:SNR-NumLayers-3D}, we draw a typical relative SNR gain factor as a function of ITM and ETM number of coating layers doublets defined as follows:
\begin{equation}
  G_{\rm rel}(N_{\rm ITM}, N_{\rm ETM})
  = \frac{\rho^2(N_{\rm ITM}, N_{\rm ETM})}
      {\rho^2(N_{\rm ITM}^{\rm def}, N_{\rm ETM}^{\rm def})} \,,
\end{equation}
where
\begin{align}\label{N_def}
  N_{\rm ITM}^{\rm def} &= 8\,, & N_{\rm ETM}^{\rm def} &= 19
\end{align}
are the default values for the number of coating layers doublets prescribed by GWINC for the Advanced LIGO. Cyan dots correspond to $N_{\rm ITM} = N_{\rm ITM}^{\rm def}$, and the magenta ones --- to $N_{\rm ETM} = N_{\rm ETM}^{\rm def}$. It is easy to see that for each $N_{\rm ITM}$, the optimal value of $N_{\rm ETM}$ exists, which corresponds to the balance of the quantum noise and the coating thermal noise. For smaller $N_{\rm ETM}$, the ETM transmittance is too big and the quantum noise dominates. Otherwise, the redundant number of the coating layers leads to a domination of the coating thermal noise.

At the same time, the dependence of $G_{\rm rel}(N_{\rm ITM}, N_{\rm ETM})$ on $N_{\rm ITM}$ displays no extremal behavior down to $N_{\rm ITM}=0$. This means that our optimisation algorithm tries to remove the noisy input mirrors completely, switching to the GEO-600 pure Michelson configuration \cite{Hild2006}, and providing the necessary bandwidth of the interferometer by means of the signal recycling mirror. Unfortunately, in fact, this regime can not be considered as the optimal one due to the following reasons. First, in our consideration, we have not taken into account the noise introduced by the beamsplitter. In a standard Fabry-P\'erot/Michelson configuration with arm-cavity finesse $\mathcal{F}\gg1$, the beamsplitter-induced noise can be neglected, as each reflection of the pump laser light from the beamsplitter corresponds to $\sim\mathcal{F}$ reflections inside the arm cavities, and thus it is suppressed by the factor $\mathcal{F}^{-1}$. However, it becomes important if $N_{\rm ITM}\to0$ and, therefore, $\mathcal{F}\to1$.

In particular, in the Advanced LIGO, the motion of the beamsplitter relative to the ETMs will be monitored by means of an additional modulation sidebands of the carrier light resonant only in the power and signal recycling cavities, whereupon a sufficiently small value of $\mathcal{F}$ will imply much more stringent requirements to this monitoring system \cite{T1000298}.

Second, for a given value of the circulating optical power in the arm cavities $I_c$, the power on the beamsplitter $I_{\rm BS}$ is proportional to $I_c\mathcal{F}^{-1}$, and very high power on the beamsplitter and ITMs can create undesirable effects like thermal lensing due to heating of the input mirrors and the beamsplitter by the absorbed optical power.

Due to these reasons, we limited the minimal number of the coating layers doublets on ITM by $N_{\rm ITM}=4$, which corresponds to $\mathcal{F}\approx20$. This value is sufficiently high to neglect with good precision the noise of the beamsplitter and to provide a reasonably low optical power on the beamsplitter (in particular, in all our estimates, we limited the power absorbed in each of the input mirrors and in the beamsplitter by the value of $1\,{\rm W}$). A bold red dot in Fig.\,\ref{fig:SNR-NumLayers-3D} corresponds to the optimal configuration conditional on the above mentioned constraints.

Another technical problem that may arise in the real interferometer is related to the pumping light leaving the arm cavities (that are usually impedance-matched for this light) through the ETMs. This light has a nonzero chance to return back into the cavities after scattering on the surrounding seismically not isolated objects and thus carrying random phase that may be a source of additional noise for the interferometer \cite{Evans_private}. This scattering can be prevented by placing an absorbing plates with the power reflectivity $R_{\rm abs}$ behind the ETMs. Simple estimate shows that in the case of the specular (mirror-like) reflectivity of these plates, their positions have to be controlled with the precision about $T_{\rm ETM}\sqrt{R_{\rm abs}} \sim 10^{-4} \text{-} 10^{-5}$ times relaxed compared with the signal displacement, where $T_{\rm ETM}$ is the ETMs power transmissivity (assuming $R_{\rm abs}\approx0.01$ and $T_{\rm ETM} \sim 10^{-3} \text{-} 10^{-4}$, see Tables \ref{table:p-BNS}, \ref{table:p-Burst}). Note that in the double-mirror topology of \cite{05a1Kh}, this factor is equal to just $T_{\rm ETM}$, that is, bigger by at least one order of magnitude.

\section{Discussion}\label{sec:conclusion}

\begin{table*}
  \caption{BNS optimisation.}\label{table:gain-BNS}
  \begin{ruledtabular}
    \begin{tabular}{lcccccccc}
      Configuration & $I_c$ & $N_{\rm ITM}$ & $N_{\rm ETM}$ & $I_0\,[{\rm kW}]$ & $I_{\rm BS}\,[{\rm kW}]$  & $e^{2r}$ & $G_{\rm rel}$ & $G_{\rm abs}$  \\
      \hline
      Plain           & $200\,{\rm kW}$ & 4 & 11 & 0.5  & 16 & --  & 1.31  & 1.24  \\
      Squeezed        & $200\,{\rm kW}$ & 4 & 11 & 0.5  & 16 & 1.5 & 1.31  & 1.25  \\
      Pre-filtering   & $200\,{\rm kW}$ & 4 & 11 & 0.5  & 16 & 10  & 1.33  & 2.13  \\
      Post-filtering  & $200\,{\rm kW}$ & 4 & 12 & 0.25 & 16 & 8.7 & 1.33  & 2.03  \\
      Plain           & $840\,{\rm kW}$ & 4 & 14 & 0.4  & 68 & --  & \multicolumn{2}{c}{1.26}  \\
      Squeezed        & $840\,{\rm kW}$ & 4 & 14 & 0.4  & 68 & 1.6 & 1.26  & 1.26  \\
      Pre-filtering   & $840\,{\rm kW}$ & 4 & 14 & 0.4  & 68 & 10  & 1.30  & 2.30  \\
      Post-filtering  & $840\,{\rm kW}$ & 4 & 14 & 0.4  & 68 & 10  & 1.30  & 2.25
    \end{tabular}
  \end{ruledtabular}
\end{table*}

\begin{table*}
  \caption{Bursts optimisation.}\label{table:gain-Burst}
  \begin{ruledtabular}
    \begin{tabular}{lcccccccc}
      Configuration & $I_c$ & $N_{\rm ITM}$ & $N_{\rm ETM}$ & $I_0\,[{\rm kW}]$ & $I_{\rm BS}\,[{\rm kW}]$ & $e^{2r}$ & $G_{\rm rel}$ & $G_{\rm abs}$  \\
      \hline
      Plain           & $200\,{\rm kW}$ & 4 & 11 & 0.25 & 16 & -- & 1.24  & 0.59  \\
      Squeezed        & $200\,{\rm kW}$ & 4 & 12 & 0.25 & 16 & 10 & 1.16  & 1.22  \\
      Pre-filtering   & $200\,{\rm kW}$ & 4 & 12 & 0.25 & 16 & 10 & 1.21  & 1.41  \\
      Post-filtering  & $200\,{\rm kW}$ & 4 & 12 & 0.25 & 16 & 10 & 1.21  & 1.43  \\
      Plain           & $840\,{\rm kW}$ & 4 & 14 & 0.4  & 68 & -- & \multicolumn{2}{c}{1.20}  \\
      Squeezed        & $840\,{\rm kW}$ & 4 & 14 & 0.4  & 68 & 10  & 1.17  & 2.58  \\
      Pre-filtering   & $840\,{\rm kW}$ & 4 & 14 & 0.4  & 68 & 10  & 1.21  & 2.91  \\
      Post-filtering  & $840\,{\rm kW}$ & 4 & 14 & 0.4  & 68 & 10  & 1.21  & 2.92
    \end{tabular}
  \end{ruledtabular}
\end{table*}

The results of the optimisation are given in Tables \ref{table:gain-BNS} and \ref{table:gain-Burst} for the BNS and burst sources, respectively. In addition, the explicit values of the optimised vectors ${\bf p}$ are shown in Tables \ref{table:p-BNS} and \ref{table:p-Burst} in the Appendix.

It follows from these results, that reducing $N_{\rm ITM}$ from $N_{\rm ITM}^{\rm def}=8$ to 4, and $N_{\rm ETM}$ from $N_{\rm ETM}^{\rm def}=19$ to 14 for $I_c=840\,{\rm kW}$ and to 11-12 for $200\,{\rm kW}$, it is possible to increase the signal to noise ratio $\rho^2$ for the BNS events by $\sim30\%$, and for the burst events --- by $\sim20\%$, see the corresponding columns labeled ``$G_{\rm rel}$''. A few per cent surplus of $G_{\rm rel}$ for low circulating power case ($I_c=200\,{\rm kW}$) over the high power one ($I_c=840\,{\rm kW}$) is simply the result of a smaller number of coating layers required in the former case ($N_{\rm ETM}=11\text{-}12$ vs 14), {\it i.e.} the less circulating power is required for the same maximal input power, the more transparent can be the end mirrors.

We are also able to compare the performance of the interferometers which employ different advanced techniques like squeezing, pre- and post-filtering with the baseline setup, characterized by
\begin{align}\label{default}
  I_c &= 840\,{\rm kW}\,, & r &= 0 \,, &
  N_{\rm ITM}^{\rm def} &= 8\,, & N_{\rm ETM}^{\rm def} &= 19\,.
\end{align}
For this purpose, we introduce the absolute gain defined as:
\begin{equation}
  G_{\rm abs} = \frac{\rho^2}{\rho_{\rm base}^2} \,,
\end{equation}
with $\rho_{\rm base}^2$ related to the baseline interferometer. The corresponding values of $G_{\rm abs}$ are given in Tables \ref{table:gain-BNS} and \ref{table:gain-Burst}.

These numbers tell us that for the GW bursts, the near-four-fold increase of the circulating power results in the two-fold relative rise of the SNR, yet almost the same result can be achieved by employing a 10~dB input squeezing. At the same time, the BNS signal-to noise ratio $\rho^2_{\rm BNS}$ remains almost unchanged with the power increase, and does not benefit noticeably from the squeezing (note that the optimisation algorithm virtually rejects the squeezing, using only small value of $r$).

The physics behind such a behavior is transparent. For the sensitivity of the detector to GW bursts depend on the high frequency shot-noise-dominated part of the noise budget (inversely proportional to the product of circulating power $I_c$ and squeezing factor $e^{2r}$) to a greater extent ($h_{\rm burst}(\Omega)\propto \Omega^{-1/2}$) compared to the sensitivity to gravitational waves emitted by compact binary systems ($h_{\rm BNS}(\Omega)\propto \Omega^{-7/6}$), the former one displays the expected trend described above. At the same time, BNS sensitivity assigns more weight to the radiation-pressure-dominated low-frequency region and to medium frequencies equally sensitive to both parts of quantum noise, thus making it much harder to balance their influence, for the radiation pressure fluctuations have opposite dependence on power and squeezing factor as the shot noise.

Another apparent conclusion yielding from our optimisation is that even a relatively short ($l=50$~m) filter cavity is capable of quite significant sensitivity gain for both considered GW sources. Though there is no significant difference in performance for pre- and post-filtering schemes, given the assumed values of cavity length and losses per bounce, the pre-filtering scheme experimental implementation is easier and thus is more favorable.

\begin{figure*}
  \includegraphics[width = 0.49\textwidth]{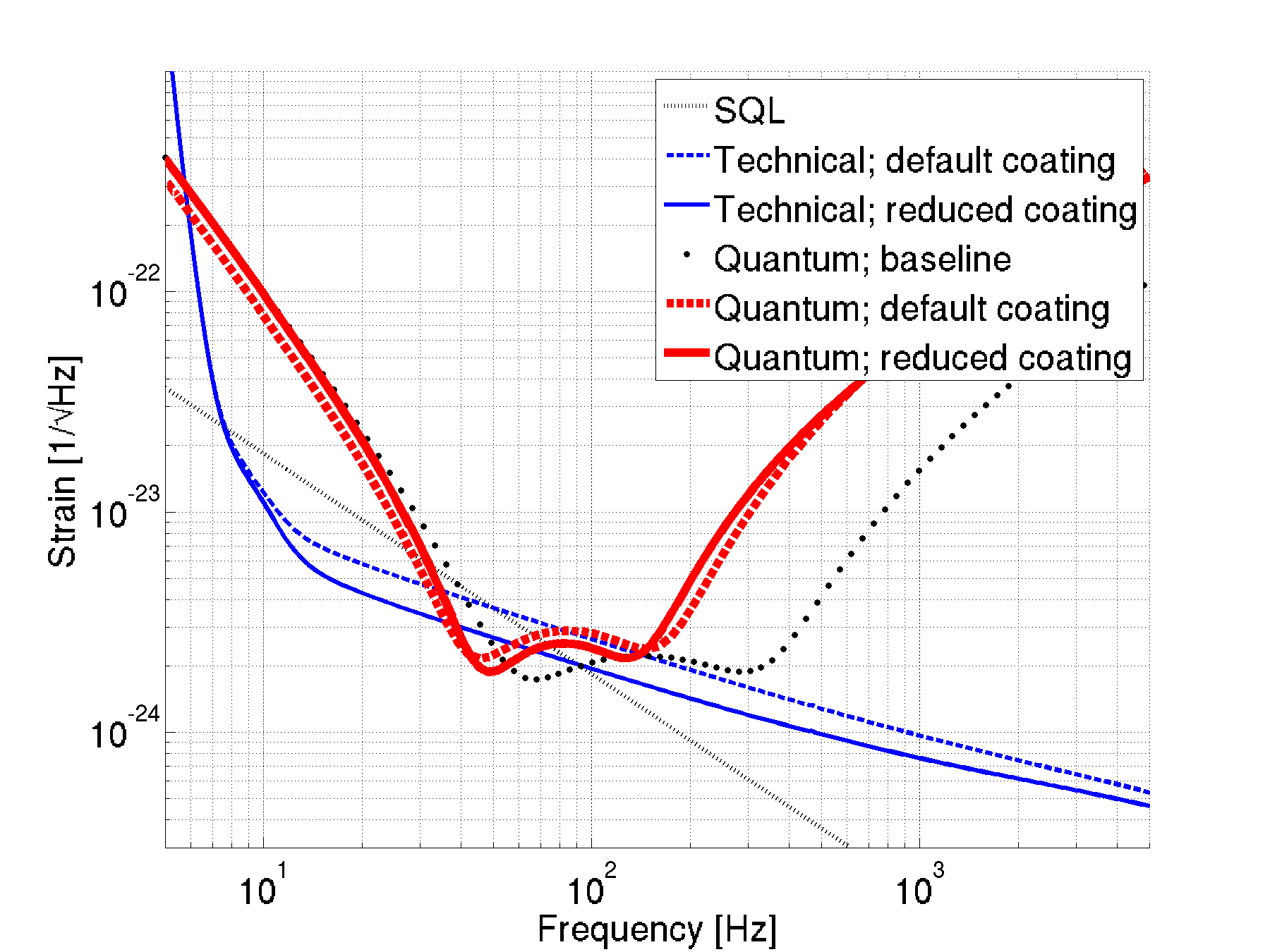}\hfill
  \includegraphics[width = 0.49\textwidth]{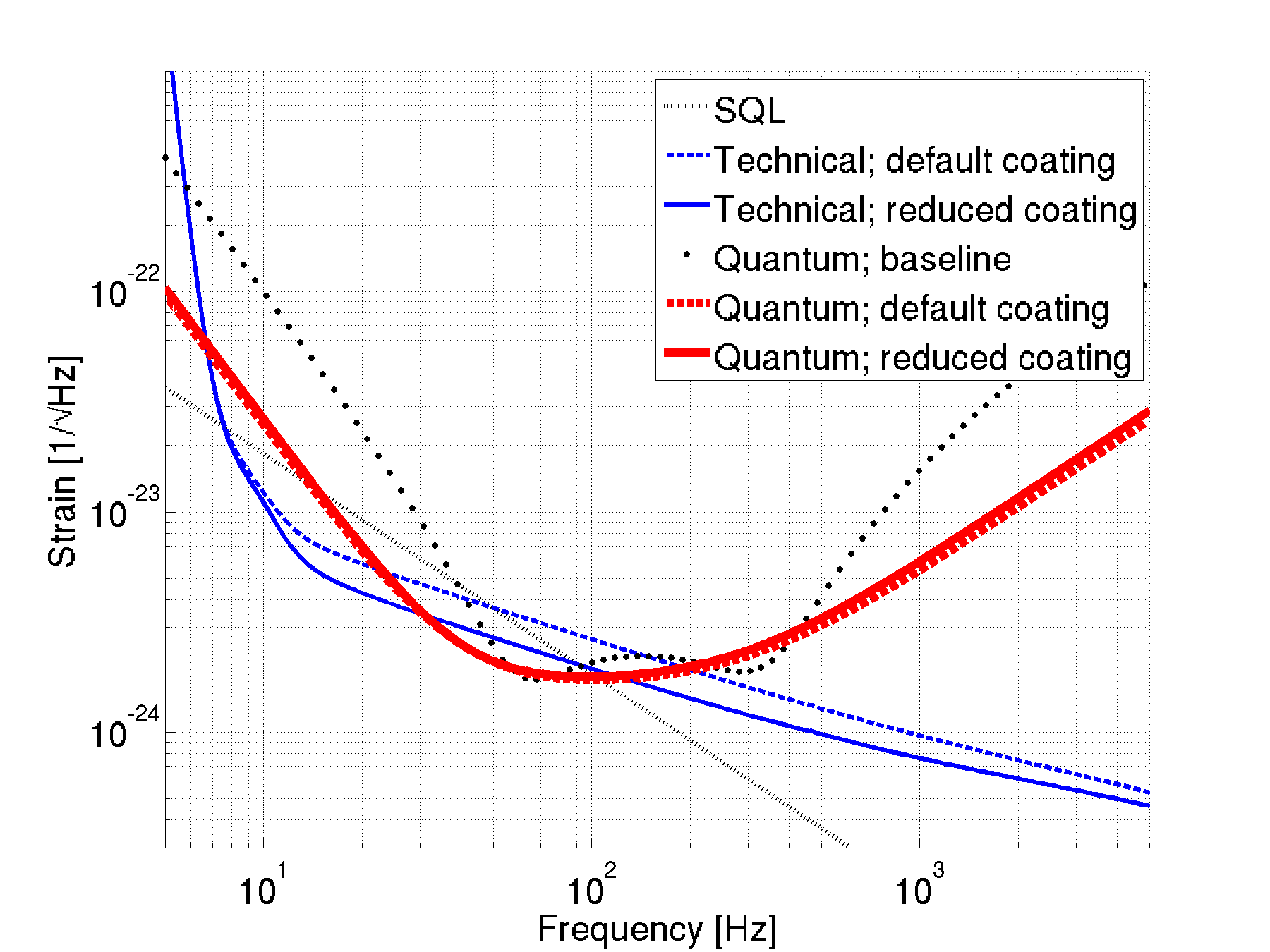}
  \includegraphics[width = 0.49\textwidth]{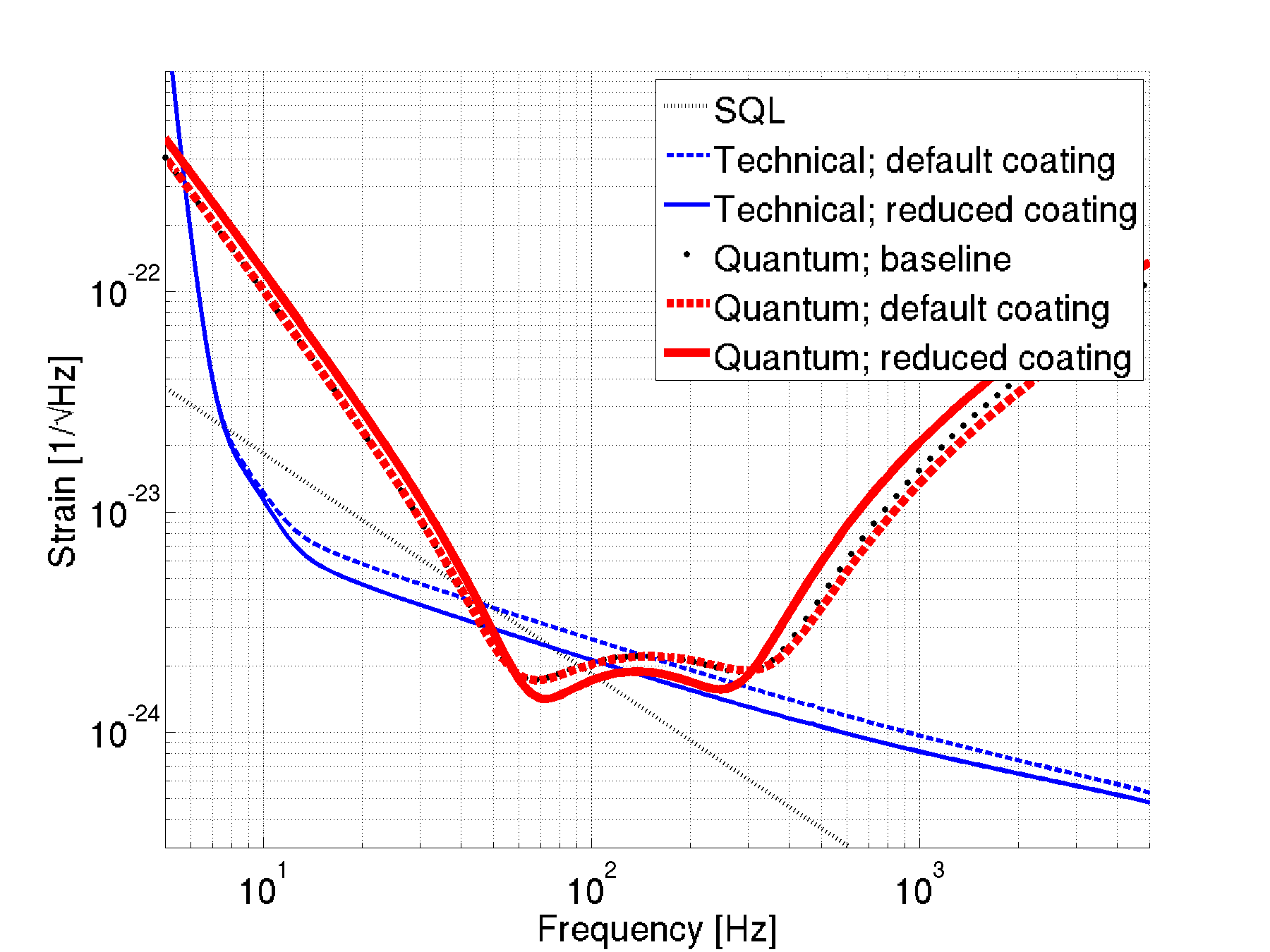}\hfill
  \includegraphics[width = 0.49\textwidth]{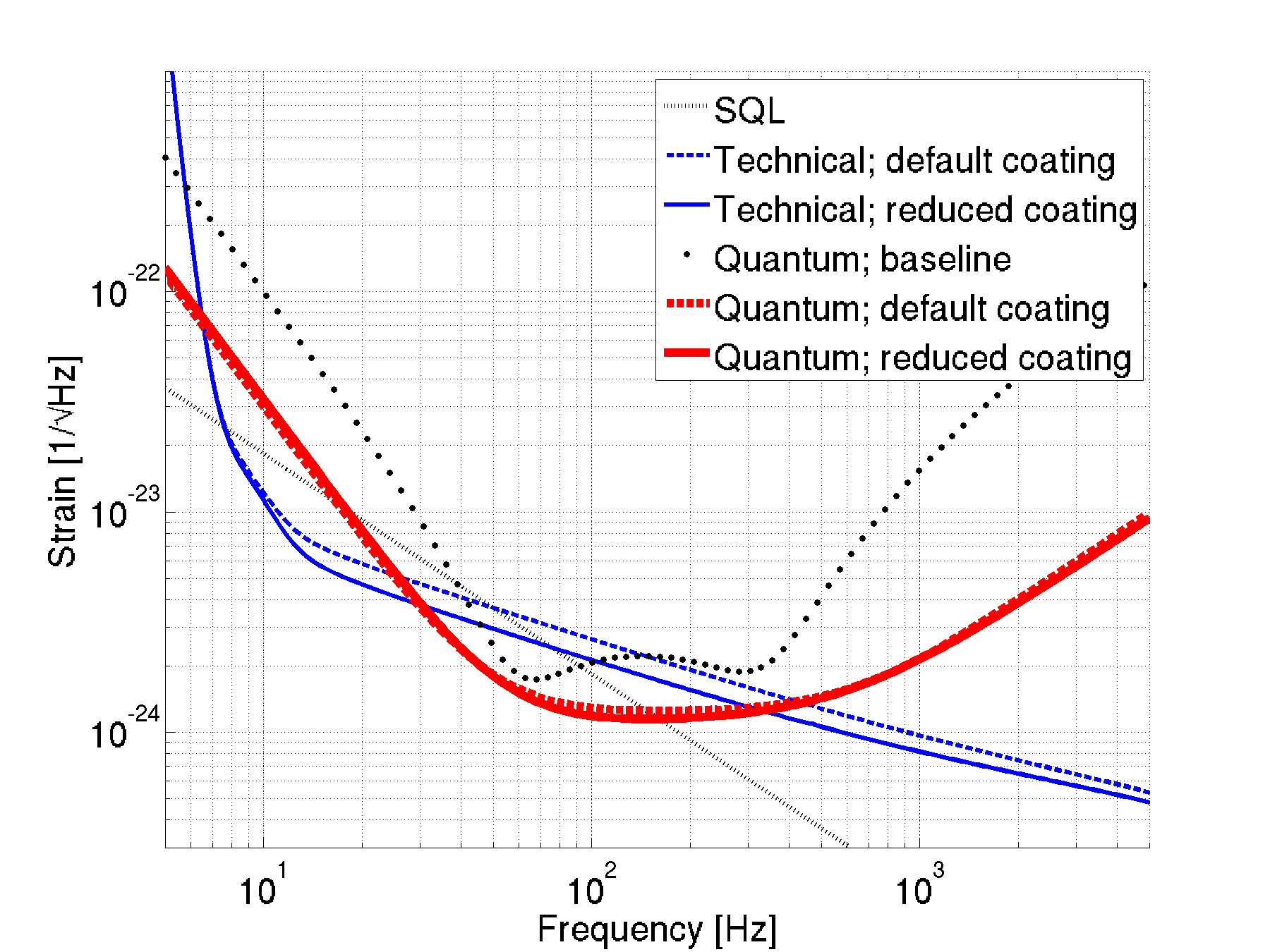}
  \caption{Optimal quantum and technical noise spectral densities for the BNS sources. Left column: schemes with frequency-independent squeezing (referred to as ``squeezed''); right column: schemes with frequency-dependent input squeezing (referred to as ``pre-filtering''). Top row: $I_c=200\,{\rm kW}$, bottom row: $I_c=840\,{\rm kW}$. Solid lines: optimised numbers of the coating layer doublets $N_{\rm ITM}$ and $N_{\rm ETM}$; dashed lines: default numbers $N_{\rm ITM}=8$ and $N_{\rm ETM}=19$. Dash-dotted line: ``baseline'' configuration: $I_c=840\,{\rm kW}$, $N_{\rm ITM}=8$ and $N_{\rm ETM}=19$, no squeezing; dotted line: SQL.}\label{fig:nsns}
\end{figure*}

\begin{figure*}
  \includegraphics[width = 0.49\textwidth]{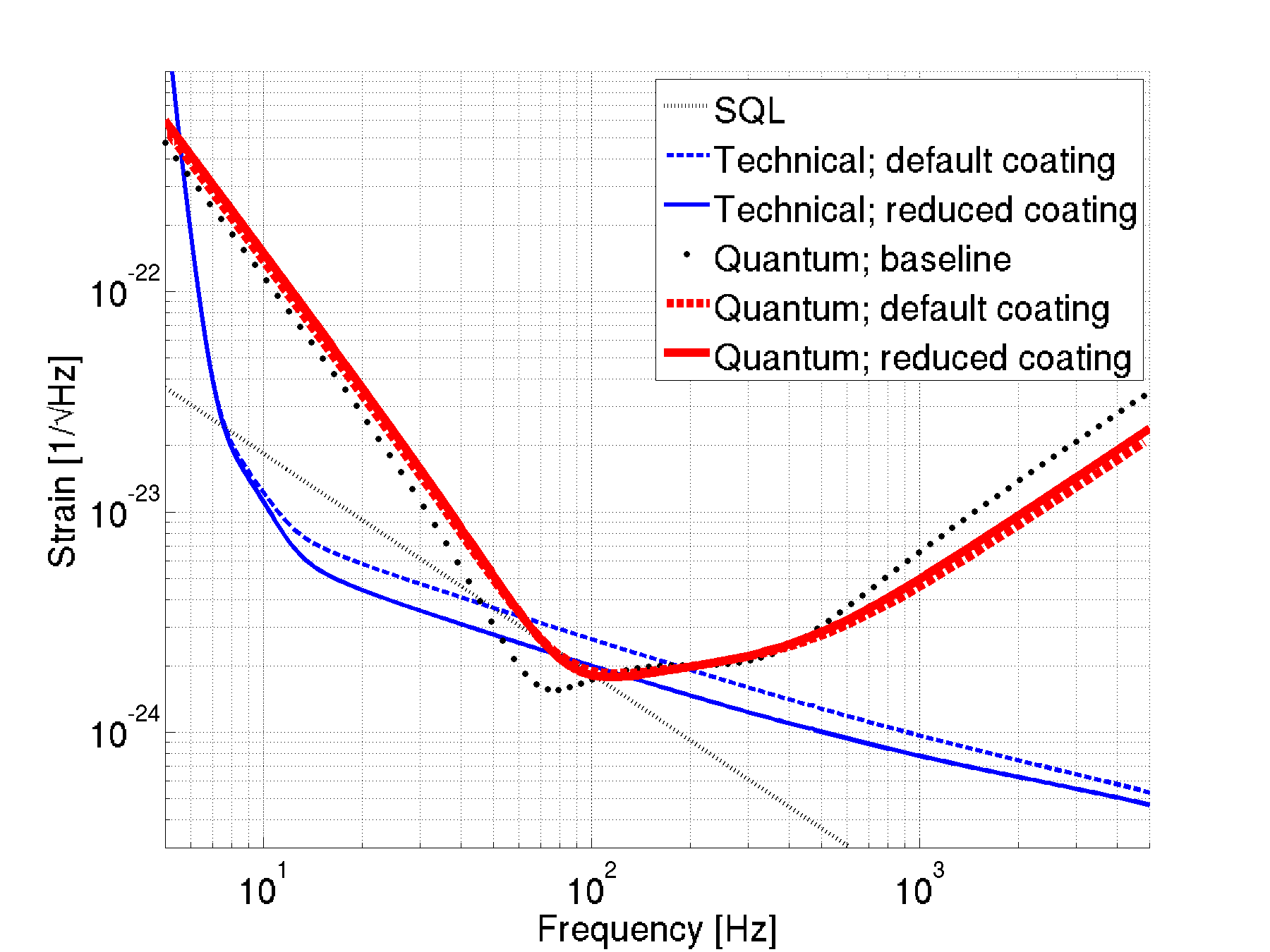}\hfill
  \includegraphics[width = 0.49\textwidth]{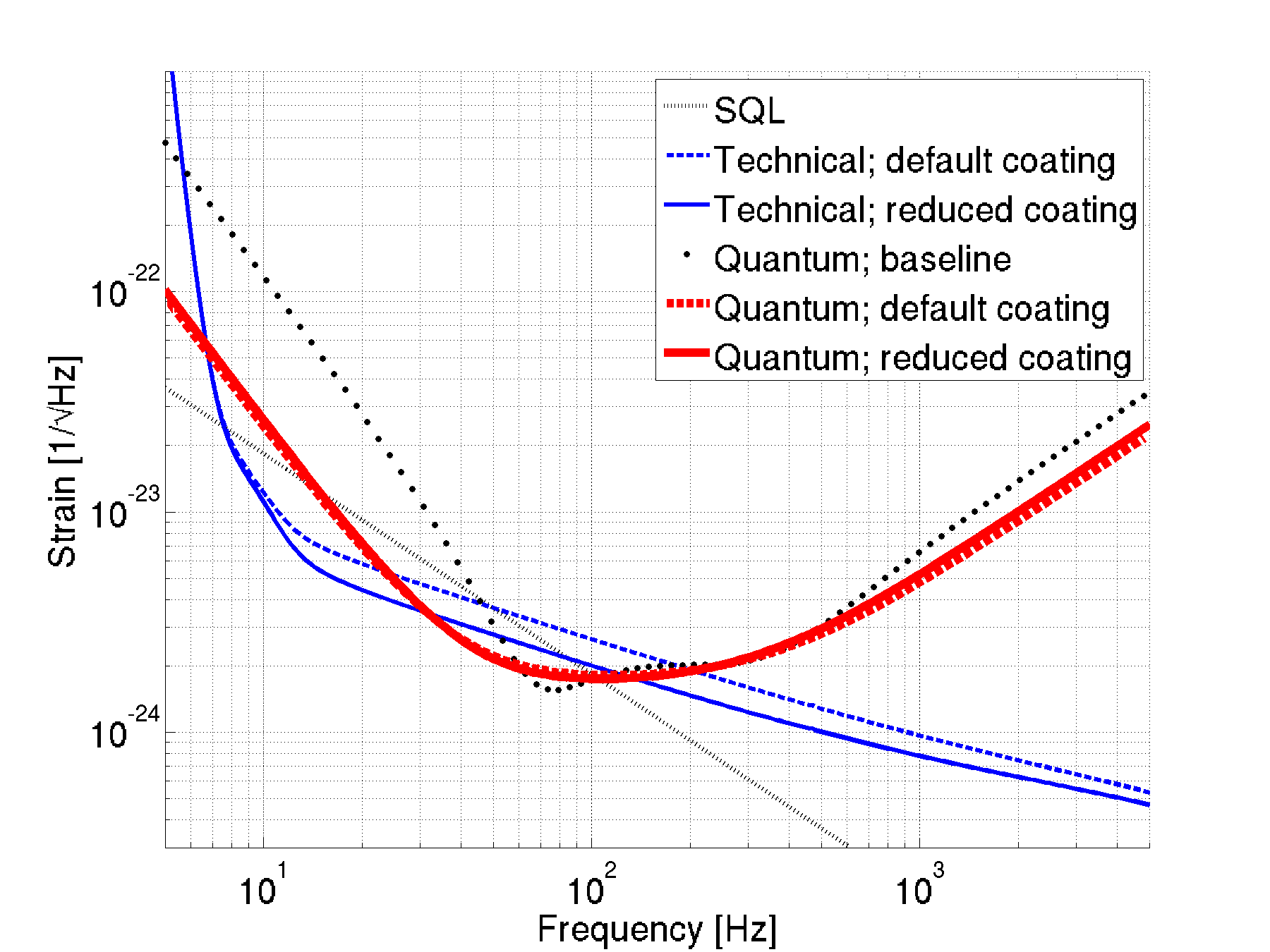}
  \includegraphics[width = 0.49\textwidth]{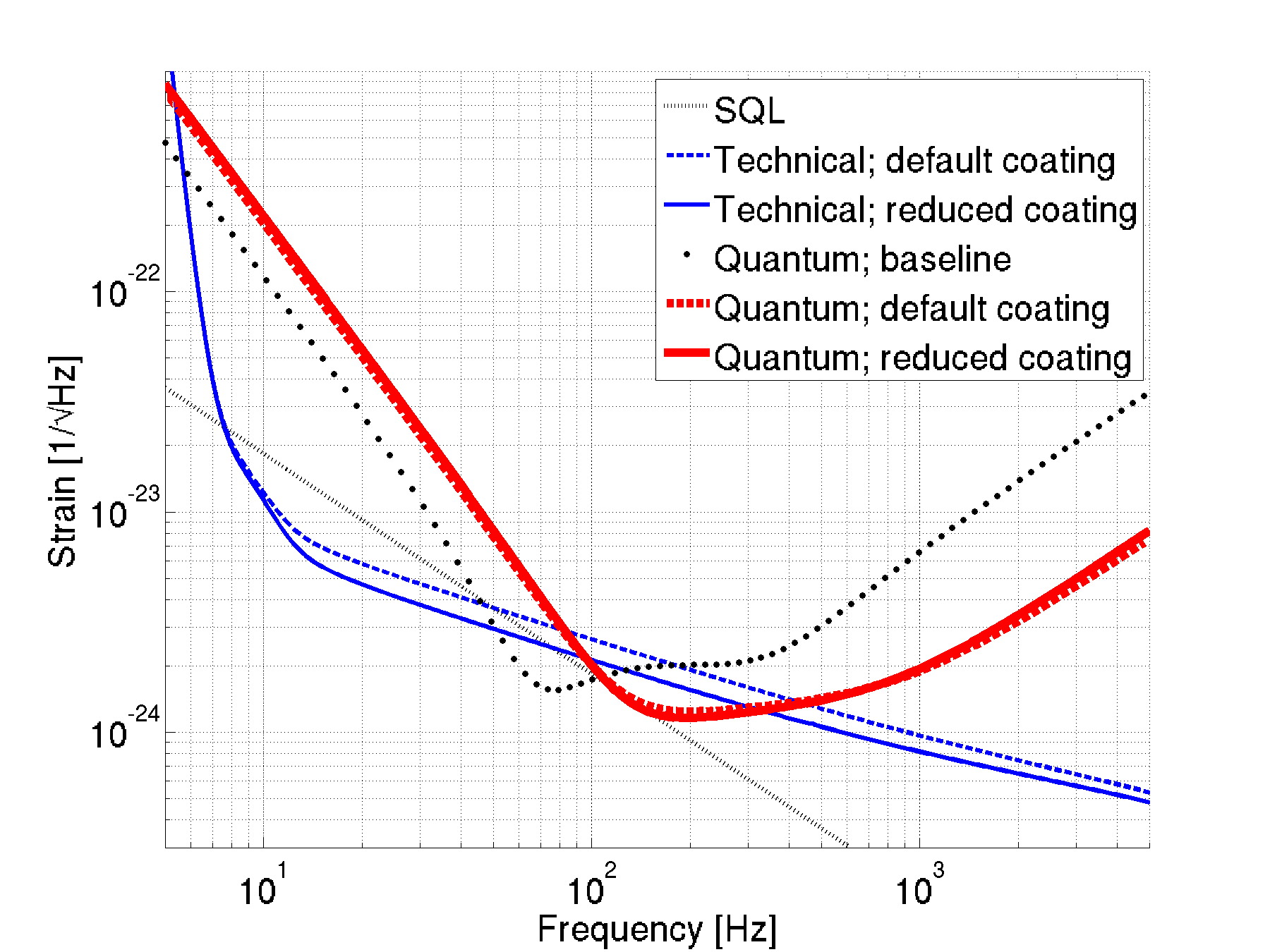}\hfill
  \includegraphics[width = 0.49\textwidth]{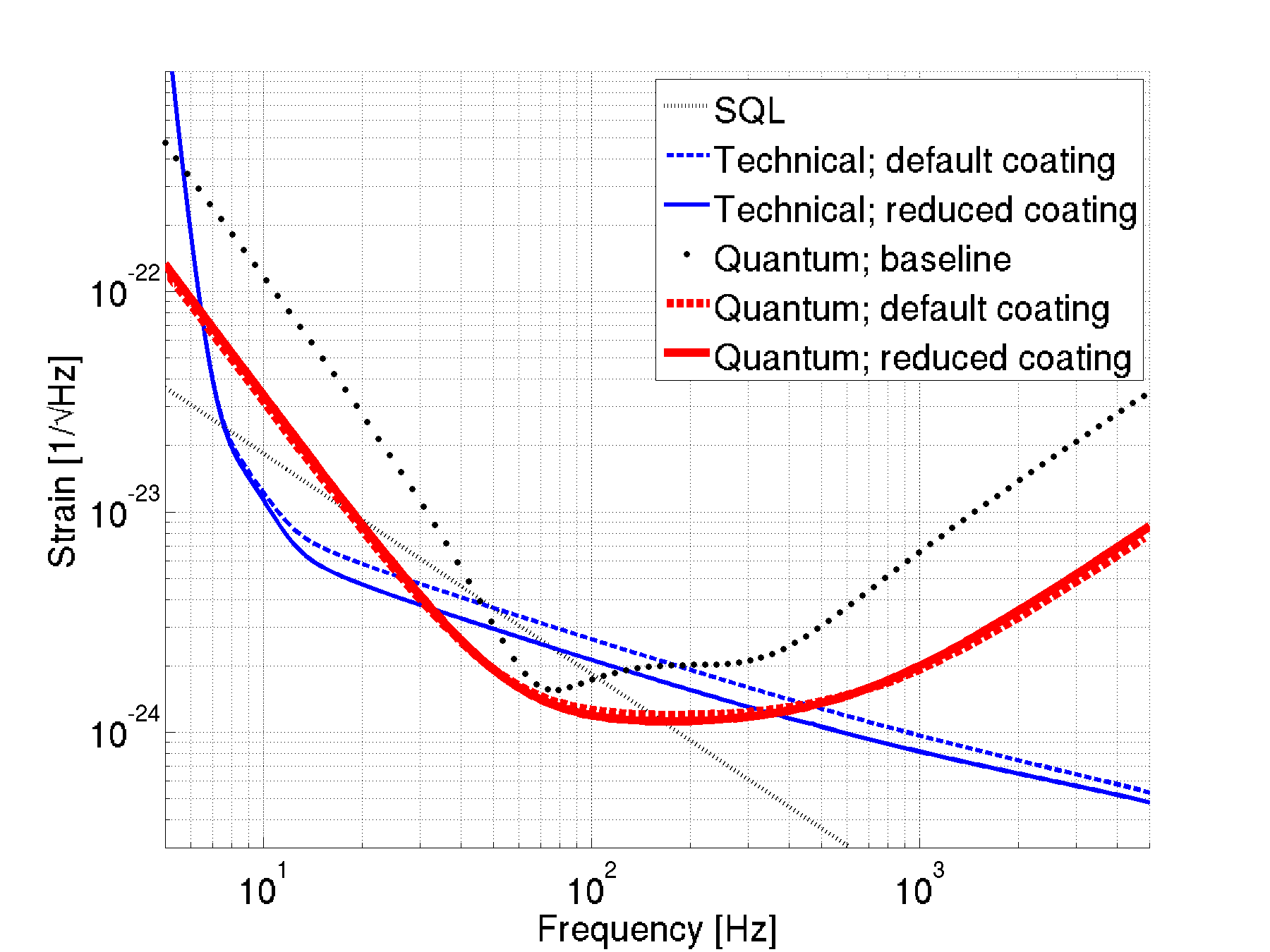}
  \caption{Optimal quantum and technical noise spectral densities for the burst sources. Left column: schemes with frequency-independent squeezing (referred to as ``squeezed''); right column: schemes with frequency-dependent input squeezing (referred to as ``pre-filtering''). Top row: $I_c=200\,{\rm kW}$, bottom row: $I_c=840\,{\rm kW}$. Solid lines: optimised numbers of the coating layer doublets $N_{\rm ITM}$ and $N_{\rm ETM}$; dashed lines: default numbers $N_{\rm ITM}=8$ and $N_{\rm ETM}=19$. Dash-dotted line: ``baseline'' configuration, see Eq.\,\eqref{default}; dotted line: SQL.}\label{fig:burst}
\end{figure*}

Plots of the optimal spectral densities for both quantum and technical noise are drawn in Figs.\,\ref{fig:nsns} and \ref{fig:burst} for BNS and GW bursts, respectively. We elected for plotting only the most promising variants with fixed and frequency-dependent input squeezing. For comparison, the corresponding spectral densities for the default numbers of the coating layers doublets \eqref{N_def}, as well as for the ``baseline'' configuration \eqref{default} are shown there as well. These plots demonstrate that the optimisation procedure considered here, while suppressing noticeably the sum technical noise, introduces only minor changes into the quantum noise. The only exception is the BNS optimisation in the absence of the filter cavity, which reduces the quantum noise in the best sensitivity area around $100\,{\rm Hz}$, following the technical noise, at the price of narrower width of this region.

We would like to emphasize that our solution to the coating thermal noise problem provides a 20-30\% gain in the signal-to-noise ratio for the two most probable astrophysical sources of GWs which amounts in the 30-50\% increase of the event rate without any significant change to the optical layout of the GW interferometer. The  price one has to pay, namely about two-fold increase of the input laser power, slightly higher absorbed power in the core optics ($\sim 1$~W), and other issues discussed above, does not look too high in view of the potential benefits.

\acknowledgments

The authors are very grateful to Matt Evans for going to the trouble of thorough reading the manuscript and, especially, for his critical remarks that were very illuminating and useful for improving the quality of this article.

This work was supported by the Russian Foundation for Basic Research Grant No. 08-02-00580-a. The work of F.Ya.Khalili was supported by NSF and Caltech Grant No.\,PHY-0967049. Stefan Danilishin was supported by the David and Barbara Groce Startup Fund at the California Institute of Technology via the CRDF Grant RUP1-32011-MO-10.

The paper has been assigned LIGO document number P1100199.

\appendix

\section{Quantum noise spectral densities}\label{app:quantum}

\subsection{Notations}

We base our consideration here on the Caves-Schumaker's two-photon formalism \cite{Caves1985, Schumaker1985}. Treatment of all of the considered schemes of the interferometers is done is accordance with \cite{12a1DaKh} where all the detailed derivations can be found. Two-photon quadrature vectors are denoted by boldface letters, and their components --- by upright letters, {\it e.g.}
\begin{equation}
  \mathbf{\hat{a}} = \binom{\hat{{\rm a}}_1}{\hat{{\rm a}}_2} .
\end{equation}
$2\times2$ matrices are denoted by ``blackboard bold'' letters, {\it e.g.} $\mathbb{R}$.

The quadrature amplitudes are normalized in such a way, that in the vacuum quantum state, their single-sided spectral densities are equal to one, {\it e.g.}
\begin{equation}
  S[\hat{{\rm g}}_1] = S[\hat{{\rm g}}_2] = 1 \,.
\end{equation}

We use the following notation for the norms of the two-components quadrature vectors:
\begin{equation}
  \forall {\bf A}=\binom{A_1}{A_2}\,:\quad \|{\bf A}\|^2 = {\bf A}^{\sf T}{\bf A} \,.
\end{equation}

\subsection{Input-output relations}

\subsubsection{Michelson/Fabry-P\'erot interferometer}

Following the treatment of paper \cite{Buonanno2003} (often referred to as ``scaling law''), the Fourier-domain input/output relations of a signal-recycled Michelson/Fabry-P\'erot interferometer detector can be written in the equivalent form of the single effective Fabry-P\'erot cavity input/output relations:
\begin{equation}\label{eq:IO-relation}
  \mathbf{\hat{b}}(\Omega) = \mathbb{R}(\Omega)\mathbf{\hat{a}}(\Omega) + \mathbb{T}(\Omega)\mathbf{\hat{g}}(\Omega) + \mathbf{R}(\Omega)\chi(\Omega)\frac{G(\Omega)}{2}.
\end{equation}
Here $\hat{{\bf a}}$, $\hat{{\bf b}}$, $\hat{{\bf g}}$ are the quadrature vectors of, respectively, the input light, the output light and the vacuum noise which arises due to the optical losses in the interferometer,
\begin{subequations}
\begin{gather}
  \mathbb{R}(\Omega) =
  \frac{1}{\mathcal{D}(\Omega) - J \delta/\Omega^2}
    \begin{pmatrix}  R_{11} & R_{12} \\ R_{21} & R_{22} \end{pmatrix}, \\
  R_{11}
    = R_{22} = 2\gamma_1(\gamma - \ii\Omega) - \mathcal{D}(\Omega) + J\delta/\Omega^2\,,
  \nonumber \\
  R_{12} = - 2\gamma_1\delta \,, \quad
  R_{21} = 2\gamma_1\delta  -  2J\gamma_1/\Omega^2 \,, \nonumber \\
  \mathbb{T}(\Omega) =
  \frac{2 \sqrt{\gamma_1\gamma_2}}{\mathcal{D}(\Omega) - J \delta/\Omega^2}
  \begin{pmatrix}
    \gamma - \ii\Omega     &   - \delta        \\
    \delta  -  J/\Omega^2  &   \gamma - \ii\Omega
  \end{pmatrix},
\end{gather}
\end{subequations}
\begin{equation}
  \mathcal{D}(\Omega) = (\gamma - i\Omega)^2 + \delta^2 \,, \quad
  J = \frac{8 \omega_p I_c}{M c L} \,,
\end{equation}
\begin{subequations}
  \begin{gather}
    \delta   = \frac { 2 \sqrt{R_{\rm SRM}} \sin(2\phi_{\rm SR}) \gamma_{\rm ITM} }
                    { 1 + 2\sqrt{R_{\rm SRM}}\cos(2\phi_{\rm SR}) + R_{\rm SRM} } \,, \\
    \gamma_1 = \frac { \left(1-R_{\rm SRM}\right) \gamma_{\rm ITM} }
                    { 1 + 2\sqrt{R_{\rm SRM}}\cos(2\phi_{\rm SR}) + R_{\rm SRM} } \,, \\
    \gamma = \gamma_1 + \gamma_2 \,, \\
    \gamma_{\rm ITM} = \frac{c \, T_{\rm ITM}}{4L}, \quad
    \gamma_2 =\gamma_{\rm ETM} = \frac{cT_{\rm ETM}}{4L} \,,
  \end{gather}
\end{subequations}
$T_{\rm ITM}$ is the power transmissivity of the ITMs, $T_{\rm ETM}$ is the ETMs transmissivity accumulating all the optical losses in the interferometer, $R_{\rm SRM} = 1 - T_{\rm SRM}$ is the power reflectivity of SRM, and $\phi_{\rm SR} = \omega_p l_{\rm SR} / c$ is the single trip detuning phase of the carrier light in the SR-cavity with length $l_{\rm SR}$.

Vector
\begin{equation}
  \mathbf{R}(\Omega) =
  \sqrt{ \frac{2 \gamma_1 J M}{\hbar} } \frac{1}{\mathcal{D}(\Omega)}
  \begin{pmatrix}
    - \delta               \\
    \gamma - \ii\Omega
  \end{pmatrix},
\end{equation}
stands for the interferometer optical response function to the differential mechanical motion of the test masses  (dARM mode) induced by the GW tidal force:
\begin{equation}
  G(\Omega) = -ML\Omega^2h(\Omega)
\end{equation}
acting on the ETMs, and
\begin{equation}
  \chi(\Omega) = \frac{1}{-M\Omega^2 + K(\Omega)}
\end{equation}
is the mechanical susceptibility of the dARM mode modified by the optical rigidity
\begin{equation}
  K(\Omega) = \frac{M J \delta}{\mathcal{D}(\Omega)} .
\end{equation}

\subsubsection{Filter cavity}

The input/output relations for the filter cavity have the following form:
\begin{equation}\label{filter_io}
  \mathbf{\hat{o}}(\Omega) = \mathbb{R}_f(\Omega)\mathbf{\hat{i}}(\Omega) + \mathbb{T}_f(\Omega)\mathbf{\hat{q}}(\Omega),
\end{equation}
with $\mathbf{\hat{i}}$, $\mathbf{\hat{o}}$, and $\mathbf{\hat{q}}$ being the two-photon quadratures vectors of, respectively, the input, output and additional vacuum, induced by the optical losses in the filter cavity, quantum fields that fully describe all the filter cavity inputs and outputs, and
\begin{subequations}
\begin{gather}
  \mathbb{R}_f(\Omega) =
  \frac{1}{\mathcal{D}_f(\Omega)}
  \begin{pmatrix}
    R_{f 11} & R_{f 12} \\  R_{f 21} & R_{f 22}
  \end{pmatrix},\\
  R_{f 11} = R_{f 22} = \gamma_{f1}^2 - \gamma_{f2}^2 - \delta_f^2 + \Omega^2 + 2\ii\Omega\gamma_{f2}, \nonumber \\
  R_{f 12} = - R_{f 21} = - 2 \gamma_{f1} \delta_f, \nonumber   \\
  \mathbb{T}_f(\Omega) =
  \frac{2 \sqrt{\gamma_{f1}\gamma_{f2}}}{\mathcal{D}_f(\Omega)}
  \begin{pmatrix}
    \gamma_f - \ii\Omega       &
    - \delta_f               \\
    \delta_f                 &
    \gamma_f - \ii\Omega
  \end{pmatrix},
\end{gather}
\end{subequations}
\begin{equation}
  \mathcal{D}_f(\Omega) = (\gamma_f - i\Omega)^2 + \delta_f^2 \,, \\
\end{equation}
$\delta_f$ is the filter cavity detuning, $\gamma_f = \gamma_{f1} + \gamma_{f2}$ is its half-bandwidth,
\begin{equation}
  \gamma_{f1} = \frac{c \, T_f}{4L_f}, \quad \gamma_{f2} = \frac{c \, A_f}{4L_f},
\end{equation}
$T_f$ is the power transmissivity of the input/output mirror of filter cavity with length $L_f$, and $A_f$ is the coefficient of optical power losses per bounce.

\subsubsection{Homodyne detector}

We model the homodyne detector quantum efficiency $\eta<1$ by an imaginary gray filter with a power transmissivity $\eta$:
\begin{equation}\label{detector}
  \hat{{\bf d}}' = \sqrt{\eta}\,\hat{{\bf d}} + \sqrt{1-\eta}\,\hat{{\bf n}} \,,
\end{equation}
where $\hat{{\bf d}}$ is the quadrature vector of the photodetector incident field, $\hat{{\bf d}}'$ is the effective incoming field, and $\hat{{\bf n}}$ is the additional vacuum noise associated with the the photodetector quantum inefficiency $1-\eta$.

The output signal of the homodyne detector (the photocurrent) is proportional to
\begin{equation}
  i(\Omega) \propto {\bf H}^{\sf T}\hat{{\bf d}}'(\Omega)
  \propto
    {\bf H}^{\sf T}\bigl[\hat{{\bf d}}(\Omega) + \epsilon_d\hat{{\bf n}}(\Omega)\bigr],
\end{equation}
where
\begin{equation}
  \epsilon_d = \sqrt{\frac{1}{\eta} - 1}
\end{equation}
and
\begin{equation}
  {\bf H} = \binom{\cos\phi_{\rm LO}}{\sin\phi_{\rm LO}}
\end{equation}
is the homodyne vector.

\subsection{Configurations}

\subsubsection{``Plain'' interferometer}

In this case,
\begin{equation}\label{plain_out}
  \hat{{\bf d}} = \hat{{\bf b}} \,,
\end{equation}
and the incoming field $\hat{{\bf a}}$ is in the vacuum quantum state.

Combination of Eqs.\,(\ref{eq:IO-relation}, \ref{plain_out}, \ref{detector})  gives the following single-sided spectral density of the quantum noise in the units of gravitational-wave strain $h$:
\begin{multline}\label{eq:Sh-in-simplest-case}
  S^h_{\rm plain}(\Omega) = \frac{8}{M^2 L^2 \Omega^4} \\
    \times\frac{
      \|\mathbb{R}^\dag(\Omega)\mathbf{H}\|^2 + \|\mathbb{T}^\dag(\Omega)\mathbf{H}\|^2
      + \epsilon_d^2
    }
    {\abs{\mathbf{H}^{\sf T}\mathbf{R}(\Omega)\chi(\Omega)}^2} \,.
\end{multline}

\subsubsection{Squeezed input}

In this case, the incident field $\hat{{\bf a}}$ is the result of squeezing of some vacuum filed $\hat{{\bf z}}$:
\begin{equation}\label{sqz_in}
  \hat{{\bf a}}(\Omega) = \mathbb{S}\hat{{\bf z}}(\Omega) \,,
\end{equation}
with squeezing matrix defined as:
\begin{equation}
  \mathbb{S} =
  \begin{pmatrix}
    \cosh r  +  \cos 2\lambda  \sinh r   &   \sin 2\lambda  \sinh r               \\
    \sin 2\lambda  \sinh r               &   \cosh r  -  \cos 2\lambda  \sinh r
  \end{pmatrix} .
\end{equation}
Combining Eq.\,\eqref{sqz_in} with Eqs.~(\ref{eq:IO-relation}, \ref{plain_out}, \ref{detector}), we obtain:
\begin{multline}\label{eq:Sh-squeezed-input}
  S^h_{\rm sqz}(\Omega) = \frac{8}{M^2 L^2 \Omega^4} \\
    \times\frac{
      \|\mathbb{S}\mathbb{R}^\dag(\Omega)\mathbf{H}\|^2
      + \|\mathbb{T}^\dag(\Omega)\mathbf{H}\|^2
      + \epsilon_d^2
    }
    {\abs{\mathbf{H}^{\sf T}\mathbf{R}(\Omega)\chi(\Omega)}^2} \,.
\end{multline}

\subsubsection{Pre-filtering}

In the pre-filtering scheme, the incident field with frequency-dependent squeezing angle $\lambda(\Omega)$ is created by means of passing the squeezed light with frequency-independent squeezing \eqref{sqz_in} through the filter cavity, see Eq.\,\eqref{filter_io}:
\begin{equation}\label{filter_in}
  \mathbf{\hat{a}}(\Omega)
  = \mathbb{R}_f(\Omega)\mathbb{S}\mathbf{\hat{z}}(\Omega)
    + \mathbb{T}_f(\Omega)\mathbf{\hat{q}}(\Omega) \,.
\end{equation}
Combining Eq.\,\eqref{filter_in} with (\ref{eq:IO-relation}, \ref{plain_out}, \ref{detector}), we obtain:
\begin{multline}\label{eq:Sh-sqz-inputFC}
  S^h_{\rm pre}(\Omega) = \frac{8}{M^2 L^2 \Omega^4}
    \times\frac{1}{\abs{ \mathbf{H}^{\sf T}\mathbf{R}(\Omega)\chi(\Omega) }^2} \\ \times
    \Bigl[
      \|\mathbb{S}\mathbb{R}_f^\dag(\Omega)\mathbb{R}^\dag(\Omega)\mathbf{H}\|^2
      + \|\mathbb{T}_f^\dag(\Omega)\mathbb{R}^\dag(\Omega)\mathbf{H}\|^2 \\
      + \|\mathbb{T}^\dag(\Omega)\mathbf{H}\|^2
      + \epsilon_d^2
    \Bigr] .
\end{multline}

\subsubsection{Post-filtering}

In the post-filtering scheme, the outgoing field of the interferometer passes though the filter cavity:
\begin{equation}\label{filter_out}
  \mathbf{\hat{d}}(\Omega)
  = \mathbb{R}_f(\Omega)\mathbb{S}\mathbf{\hat{b}}(\Omega)
    + \mathbb{T}_f(\Omega)\mathbf{\hat{q}}(\Omega) \,.
\end{equation}
The chain of Eqs.\,(\ref{sqz_in}, \ref{eq:IO-relation}, \ref{filter_out}, \ref{detector}) gives the following spectral density:
\begin{multline}\label{eq:Sh-sqz-outpuFC}
  S^h_{\rm post}(\Omega) = \frac{8}{M^2 L^2 \Omega^4} \times
    \frac{1}
      {{\abs{ \mathbf{H}^{\sf T}\mathbb{R}_f(\Omega)\mathbf{R}(\Omega)\chi(\Omega) }^2}}
      \\ \times
    \Bigl[
      \|\mathbb{S}\mathbb{R}^\dag(\Omega)\mathbb{R}_f^\dag(\Omega)\mathbf{H}\|^2
      + \|\mathbb{T}^\dag(\Omega)\mathbb{R}_f^\dag(\Omega)\mathbf{H}\|^2 \\
      + \|\mathbb{T}_f^\dag(\Omega)\mathbf{H}\|^2
      + \epsilon_d^2
    \Bigr] .
\end{multline}

\begin{table*}[t]
  \caption{BNS optimisation.}\label{table:p-BNS}
  \begin{ruledtabular}
    \begin{tabular}{lcccccccccc}
      Configuration & $I_c$ & $T_{\rm ITM}$ & $T_{\rm ETM}$ & $R_{\rm SRM}$ & $\phi_{\rm SRC}\,[{\rm rad}]$ & $\phi_{\rm LO}\,[{\rm rad}]$ & $\lambda\,[{\rm rad}]$ & $e^{2r}$ & $\gamma_I\,[{\rm Hz}]$ & $\delta_f\,[{\rm Hz}]$ \\
      \hline
      Plain & 200kW  & 0.15  & $1 \cdot 10^{-3}$  & 0.78  & 0.30  & 2.64  & --  & --  & --  & --   \\
      Squeezed & 200kW  & 0.15  & $1 \cdot 10^{-3}$  & 0.70  & 0.29  & 2.61  & -0.01  & 1.5  & --  & --   \\
      Pre-filtering & 200kW  & 0.15  & $1 \cdot 10^{-3}$  & 0.07  & 0.01  & 1.29  & -0.33  & 10  & 158  & 213   \\
      Post-filtering & 200kW  & 0.15  & $6 \cdot 10^{-4}$  & 0.10  & -0.21  & 1.82  & 0.68  & 8.6  & 156  & 220   \\
      Plain & 840kW  & 0.15  & $1 \cdot 10^{-4}$  & 0.60  & 0.54  & 2.79  & --  & --  & --  & --   \\
      Squeezed & 840kW  & 0.15  & $1 \cdot 10^{-4}$  & 0.45  & 0.53  & 2.76  & 0.12  & 1.6  & --  & --   \\
      Pre-filtering & 840kW  & 0.15  & $1 \cdot 10^{-4}$  & 0.02  & 1.54  & 1.27  & -0.34  & 10  & 211  & 292   \\
      Post-filtering & 840kW  & 0.15  & $1 \cdot 10^{-4}$  & 0.06  & -1.17  & 1.77  & 0.83  & 10  & 190  & 264   \\
    \end{tabular}
  \end{ruledtabular}
\end{table*}

\begin{table*}[t]
  \caption{Bursts optimisation.}\label{table:p-Burst}
  \begin{ruledtabular}
    \begin{tabular}{lcccccccccc}
      Configuration & $I_c$ & $T_{\rm ITM}$ & $T_{\rm ETM}$ & $R_{\rm SRM}$ & $\phi_{\rm SRC}\,[{\rm rad}]$ & $\phi_{\rm LO}\,[{\rm rad}]$ & $\lambda\,[{\rm rad}]$ & $e^{2r}$ & $\gamma_I\,[{\rm Hz}]$ & $\delta_f\,[{\rm Hz}]$ \\
      \hline
      Plain & 200kW  & 0.15  & $1 \cdot 10^{-3}$  & 0.67  & 0.29  & 2.06  & --  & --  & --  & --   \\
      Squeezed & 200kW  & 0.15  & $6 \cdot 10^{-4}$  & 0.01  & 0.14  & 1.60  & -0.01  & 10  & --  & --   \\
      Pre-filtering & 200kW  & 0.15  & $6 \cdot 10^{-4}$  & 0.04  & 0.00 & 1.52  & -0.06  & 10  & 168  & 179   \\
      Post-filtering & 200kW  & 0.15  & $6 \cdot 10^{-4}$  & 0.03  & 0.00 & 1.62  & 0.04  & 10  & 173  & 176   \\
      Plain & 840kW  & 0.15  & $1 \cdot 10^{-4}$  & 0.38  & 0.53  & 2.00  & --  & --  & --  & --   \\
      Squeezed & 840kW  & 0.15  & $1 \cdot 10^{-4}$  & 0.05  & 1.52  & 1.59  & -0.01  & 10  & --  & --   \\
      Pre-filtering & 840kW  & 0.15  & $1 \cdot 10^{-4}$  & 0.02  & -1.57  & 1.57  & -0.01  & 10  & 253  & 254   \\
      Post-filtering & 840kW  & 0.15  & $1 \cdot 10^{-4}$  & 0.02  & -1.56  & 1.60  & 0.03  & 10  & 260  & 265   \\
    \end{tabular}
  \end{ruledtabular}
\end{table*}


\end{document}